\documentclass[aps,pra,showpacs,preprint]{revtex4-1}
\usepackage{amsmath}
\usepackage{graphicx}
\usepackage{dcolumn}
\usepackage{bm}
\usepackage{textcomp}
\usepackage{natbib}
\usepackage{upgreek}
\usepackage{enumerate}

\begin{document}
\title{Measurement of hyperfine structure in the $ \rm D_1 $ line of Rb}
	
\author{Durgesh Datar}
\altaffiliation{Department of Physics, Bangalore University, Bangalore  560\,056, India}
\author{Nikhil Kotibhaskar}
\affiliation{Department of Physics, Indian Institute of Science, Bangalore-560012, India}
\author{Sharath Ananthamurthy}
\altaffiliation{Department of Physics, Bangalore University, Bangalore  560\,056, India}
\author{Vasant Natarajan}
\email{vasant@physics.iisc.ernet.in}
	
\affiliation{Department of Physics, Indian Institute of Science, Bangalore-560012, India}

\begin{abstract}
We report a precise measurement of hyperfine constants in the $\rm{D_1} $ line ($ \rm {5\,P_{1/2}} $ state) of the two stable isotopes of Rb. The motivation for the work is to try and resolve discrepant values that exists in the literature. We use a technique that is different from other methods---one where the laser is not locked to a particular transition but scanned around it. This is advantageous because it overcomes frequency shifts due to servo-loop errors and other sources of noise in the experiment. The values in the two isotopes are: $ A = 120.510(26) $ MHz in $\rm {^{85}Rb}$, and $ A = 408.340(19) $ MHz in $\rm {^{87}Rb}$. These values are at variance with earlier values reported from our lab, but consistent with other published measurements.  
\end{abstract}

	\maketitle	

\thispagestyle{empty}

\section{Introduction}
Precision spectroscopy of hyperfine structure in the $\rm D $ lines of alkali atoms has been facilitated by the advent of tunable diode lasers and atomic vapor cells with high density \cite{DAN08}. In particular, Rb has been used for pioneering experiments in laser cooling and Bose-Einstein condensation \cite{MSW92,AEM95,UWN01}, using diode lasers. Many experiments in quantum optics have also been made possible because of the same advantages \cite{RWN03}. In all these kinds of experiments, the laser needs to be locked to a particular hyperfine transition. 

Rb has two isotopes: $ {\rm ^{85}Rb} $ and  $ {\rm ^{87}Rb} $. There are precise measurements of hyperfine structure in the D lines of the two isotopes reported in the literature. While the different measurements in the $\rm D_2 $ lines are consistent with each other, the one in the $\rm D_1 $ lines are discrepant. The two discrepant values are from Refs.~\cite{BGR91} and \cite{BDN04}, respectively; this suggests the need for further precise measurements. In an effort to resolve this discrepancy, a group in Australia has used a frequency comb and laser cooling for spectroscopy on the $\rm D_1 $ line of Rb \cite{MML08}. The use of laser-cooled atoms avoids errors due to saturated absorption spectroscopy used in the other two measurements. They find results consistent with that in Ref.~\cite{BGR91} and inconsistent with our work in Ref.~\cite{BDN04}.

In this work, we have repeated measurements of hyperfine structure using a different technique---one that does not involve locking the laser but rather scanning around a particular transition. This is advantageous because the lock point of the laser is not always at the peak center: due to electronics noise, acoustic noise, thermal fluctuations, and other sources of noise in the experiment. The scan axis of the diode laser for spectroscopy is calibrated using an acousto-optic modulator (AOM). The values in the two isotopes are $ A = 120.510(26) $ MHz in $ {\rm ^{85}Rb} $ and $ A = 408.340(19) $ MHz in $ {\rm ^{87}Rb} $. These values are again consistent with these in Ref.~\cite{BGR91} with comparable error bars, showing that our earlier measurements had unaccounted systematic errors.

\section{Experimental details}

\begin{figure*}[h]
	\centering
	\includegraphics[width=0.8\linewidth]{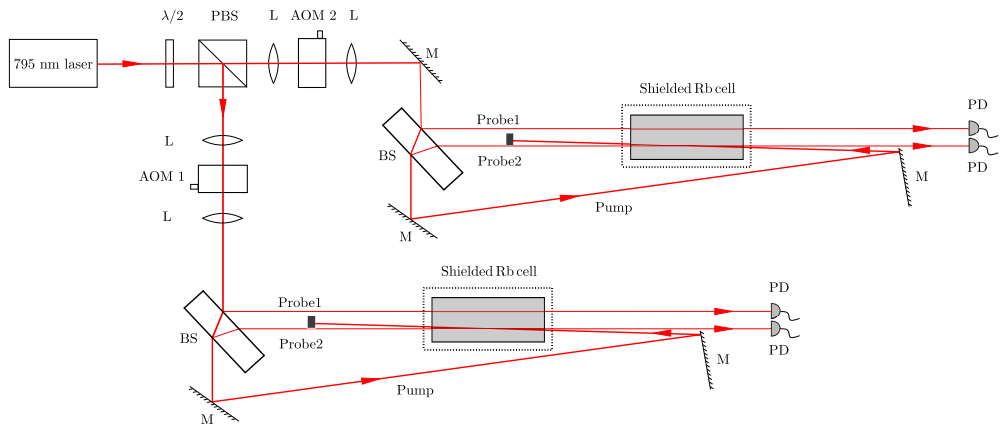}
	\caption{(Color online) Experimental schematic. Figure key: $ \lambda/2 $ -- half wave retardation plate; PBS -- polarizing beam splitter cube; L -- lens; AOM -- acousto-optic modulator; M -- mirror; BS -- beam splitter; PD -- photodiode. }
	\label{fig:rb_schematic}
\end{figure*}

The experimental setup is shown schematically in Fig.~\ref{fig:rb_schematic}. The basic idea of the measurement is to have two saturated absorption spectroscopy (SAS) systems. Each goes through an acousto-optic modulator (AOM), which for clarity are called AOM1 and AOM2 in the figure. Both SAS spectrometers use a magnetically shielded cylindrical vapor cell, of dimension 25 mm diameter $ \times $ 50 mm length. The spectra are made Doppler free by separating the signal from a second identical probe beam whose absorption is not saturated by a pump beam.

The AOMs we use work in the frequency range 300--500 MHz. Since the hyperfine interval in $ {\rm ^{85}Rb} $ is about 360 MHz, it can be accessed using a $ +1 $ order shift. However, the interval in $ {\rm ^{87}Rb} $ is about 815 MHz, hence it requires AOM1 to be adjusted for $ +1 $ order and AOM2 for $-1$ order. The RF frequency for the AOM drivers is set by a common frequency generator (HP 8656B) with a timebase accuracy of $10^{-6}$. For both isotopes, the unshifted spectrum is  obtained by not having any shift through AOM1 (zero-order beam).

The laser is a home-built grating-stabilized diode laser, as described in Ref.~\cite{MRS15}. The free running wavelength of the diode is close to 795 nm, and its total power before feedback is 150 mW. The grating used for feedback has 1800 lines/mm, and is mounted on a piezo electric transducer (PZT) so that the laser frequency can be scanned electronically. The beam coming out of the laser is Gaussian and elliptic with $ 1/e^2 $ diameter of $ 2\times 7$ mm. The probe power in each SAS spectrometer is about 100 $\upmu$W, while the pump power is $ 90 \times $ higher. Thus, the intensity at the center of the probe beam (its maximum value) is 1.82 $\rm mW/cm^2 $, which is roughly equal to the saturation intensity of 1.64 $\rm mW/cm^2 $. 

\section{Measurements in $ {\rm ^{87}Rb} $}

\begin{figure}[h]
	\centering
	\includegraphics[width=0.8\linewidth]{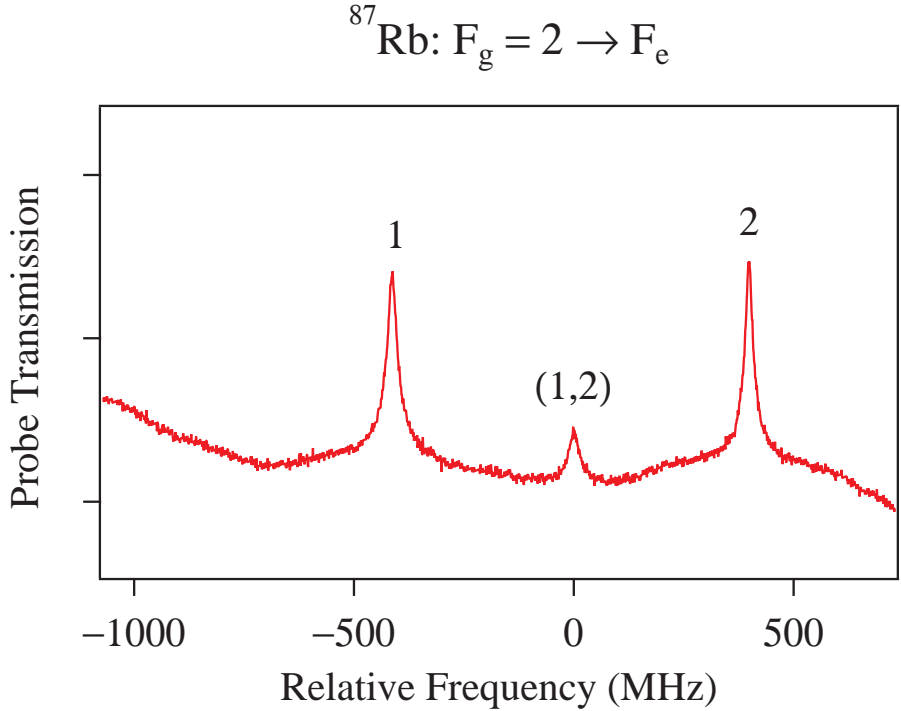}
	\caption{(Color online) Doppler-subtracted SAS spectra for $ F_g = 2 \rightarrow F_e $ transitions in $\rm {^{87}Rb} $. Each peak is labeled with the corresponding value of $ F_e $, and the crossover resonance in between with both values.}
	\label{fig:rb87_spectra}
\end{figure}

We first consider measurements in this isotope, because the discrepancy of our earlier measurement from the value in Ref.~\cite{BGR91} is quite large. The SAS spectrum obtained for $ F_g =2 \rightarrow F_e $ transitions are shown in Fig.~\ref{fig:rb87_spectra}. The linewidth of each peak is 18--20 MHz, which is larger than the 6 MHz natural linewidth, but is typical in SAS spectra and arises due to power broadening by the pump beam and a small misalignment angle between the pump/probe beams. As mentioned before, the unshifted spectrum shown is obtained by using the zero-order beam from AOM1. The interval between the $ F_e =2 $ and $ F_e = 1 $ hyperfine levels ($ \approx 815$ MHz) is measured by taking the $ +1 $ order from AOM1 and $ -1 $ order from AOM2. This will result in the interval being given by twice the AOM frequency.

The experimental method now consists varying the RF frequency driving the AOMs from 400 to 415 MHz in steps of 0.5 MHz. AOM-shifted peaks (denoted by primes) are then fitted to Lorentzian functions. The fit gives the location and error in the location, where the error depends on the signal-to-noise ratio (SNR). High SNR (and correspondingly small error) is obtained by scanning the laser only around the (1,2) crossover resonance. The two SAS spectrometers then give AOM-shifted peaks corresponding to the 2 and 1 hyperfine levels, respectively. The measured values are then fitted to a second-order polynomial, with each point being weighted by its error bar. The zero crossing of the fit gives the hyperfine interval.

\begin{figure}[h]
	\centering
	\includegraphics[width=0.8\linewidth]{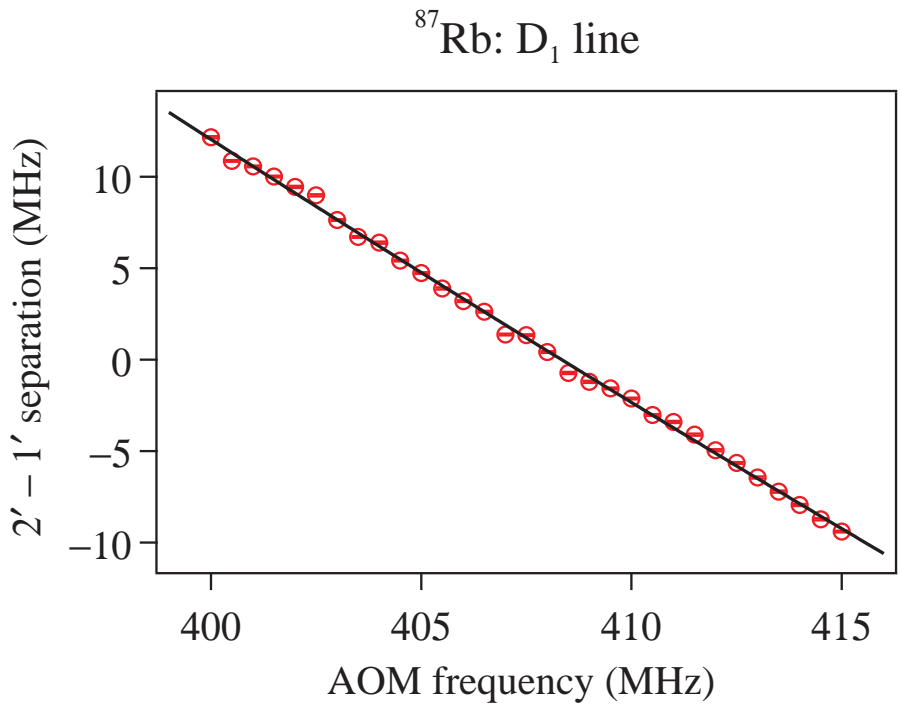}
	\caption{(Color online) Peak separation between the $ F_e = 2 $ peak and $ F_e = 1 $ AOM-shifted peaks, plotted as a function of AOM frequency. The solid line is a weighted 2nd order polynomial fit, weighted by the error bar for each point. The error bar for each point is smaller than the symbol, and not seen.}
	\label{fig:hyp87rb}
\end{figure}

The result of such a measurement in the $ \rm{5\,P_{1/2}} $ state of ${\rm ^{87}Rb} $ is shown in Fig.~\ref{fig:hyp87rb}. Note that the interval is independent of scaling of the laser scan axis, because any rescaling will only change the $ y $-axis of the figure without changing the zero crossing. We have also verified that the zero crossing of the fit remains unchanged (within its error) when we use higher-order polynomials than the second order shown in the figure---first order (linear) is not correct, because the scan axis is inherently nonlinear varying as the sine of the grating angle.

The zero crossing of the polynomial fit along with its error yields a value of the interval as $ 816.680 \pm 0.020$ MHz. The hyperfine interval is related to the hyperfine constant as $ 2A $. Therefore the measured value of the constant is $ A = 408.340 \pm 0.010 $ MHz.

\subsection{Error analysis}
The different sources of error in the measurement, and our estimated value for each, are listed below.
\begin{enumerate}
	\item Statistical error in the curve fit -- 10 kHz.
	\item AC Stark shift -- 10 kHz.
	\item Optical pumping into magnetic sublevels in the presence of stray magnetic fields (Zeeman shift) -- 10 kHz.
	\item Velocity redistribution of the atoms in the vapor cell due to radiation pressure -- 5 kHz.
	\item Collisional shifts -- 5 kHz.
	\item AOM frequency timebase error -- 0.5 kHz.
	\item Servo-loop errors in locking the laser -- 0.
\end{enumerate}
As mentioned in the introduction, the last source of error is 0 because we do not lock the laser. Adding all the other sources of error in quadrature yields the final error in the measurement as 19 kHz. 

Thus, the value of the hyperfine constant measured in this work in $\rm ^{87}Rb$ is
\[
A = 408.340 \pm 0.019 \ \ \text{MHz}
\]

\subsection{Comparison to earlier results}
\begin{figure}[h]
	\centering
	\includegraphics[width=0.8\linewidth]{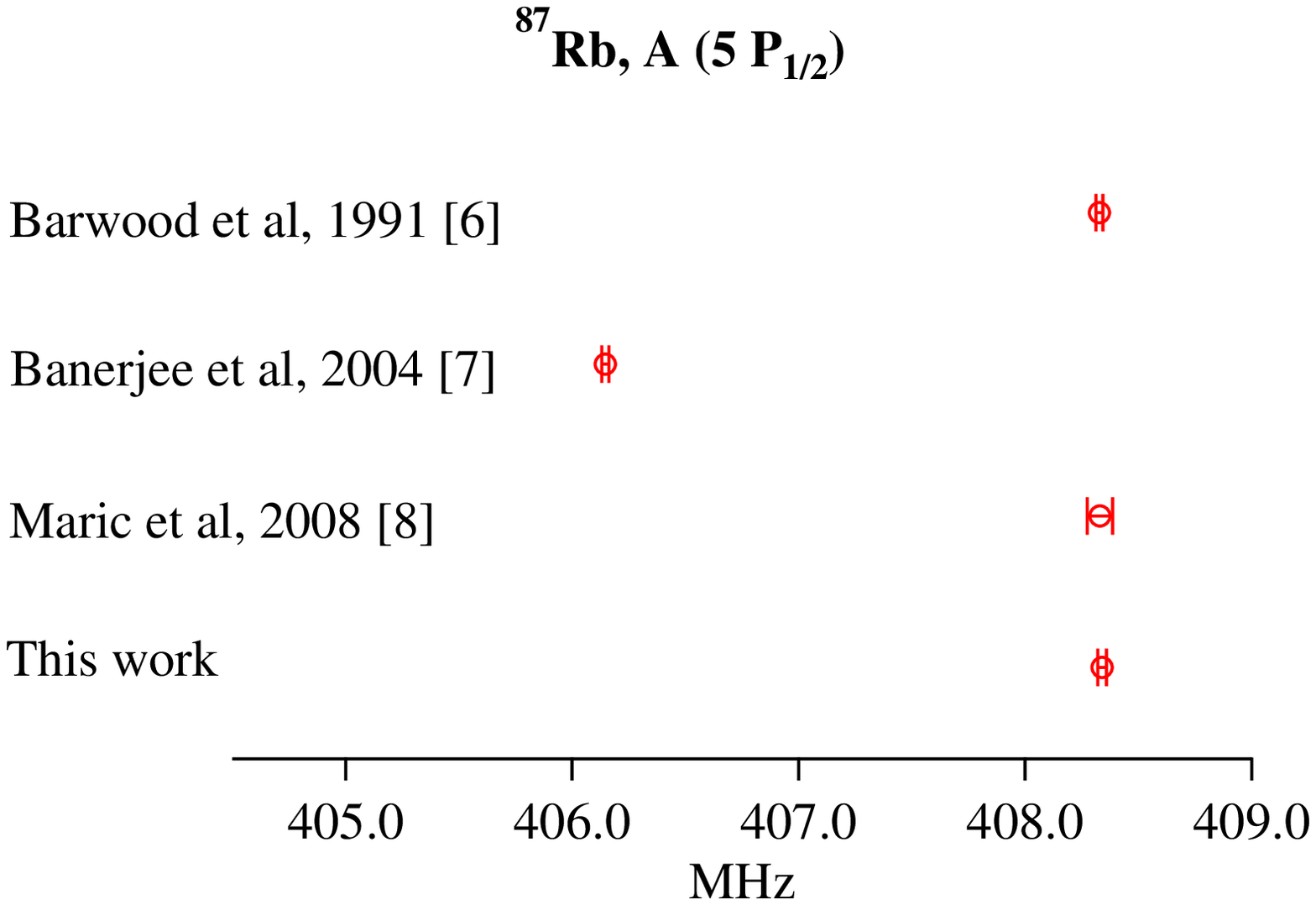}
	\caption{(Color online) Hyperfine constant $ A $ in the $ \rm 5\,P_{1/2} $ state of $ \rm {^{87}Rb} $ measured in this work compared to earlier values.}
	\label{fig:a87rb}
\end{figure}

Previous measurements of this hyperfine constant are compared in Fig.~\ref{fig:a87rb}. 

As seen, our present measurement is consistent with those in Refs.~6 and 8, but completely inconsistent with that in Ref.~\cite{BDN04}. This suggests that the measurement in Ref.~\cite{BDN04} had unaccounted systematic errors.

\section{Measurements in $ {\rm ^{85}Rb} $}

The values of the hyperfine constant in $ {\rm ^{85}Rb} $ from Refs.~6 and 7 are not that discrepant, with a difference of only $ 6 \sigma $. However, even in this case, the result from Ref.~\cite{MML08} overlaps with that of Ref.~\cite{BGR91}. We have therefore repeated measurements using the same technique for this isotope.

\begin{figure}[h]
	\centering
	\includegraphics[width=0.8\linewidth]{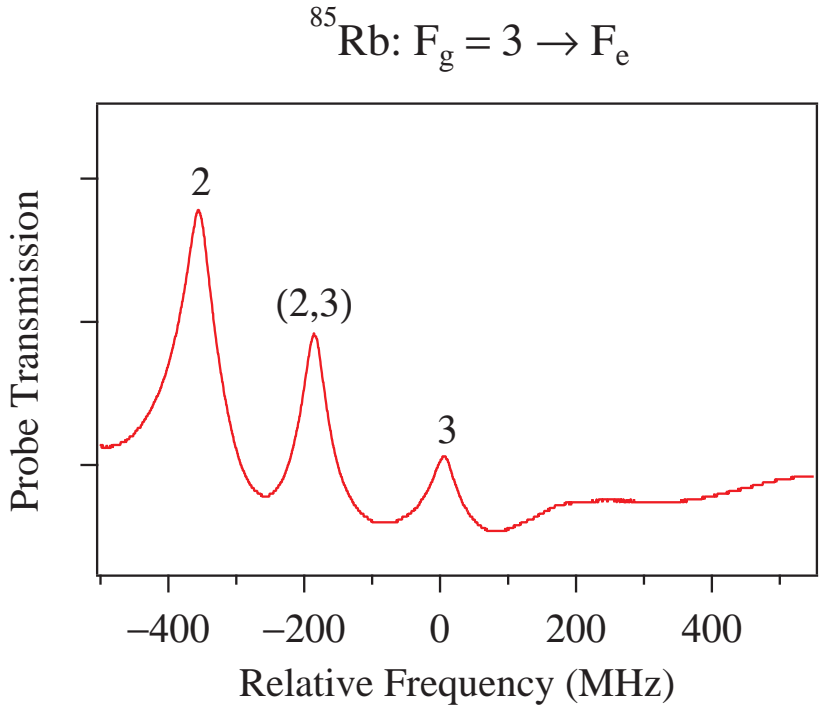}
	\caption{(Color online) Doppler-subtracted SAS spectra for $ F_g = 3 \rightarrow F_e $ transitions in $\rm {^{85}Rb} $. Each peak is labeled with the corresponding value of $ F_e $, and the crossover resonance in between with both values.}
	\label{fig:rb85_spectra}
\end{figure}

The natural abundance of $ {\rm ^{85}Rb} $ is 72\%, therefore we get much better SNR in the SAS spectrum; this is clear from the spectrum shown in Fig.~\ref{fig:rb85_spectra}. The effect of this is a smaller error in the determination of the peak center from the Lorentzian fit. The relevant separation in this case is between the $ F_e = 2 $ and $ F_e = 3 $ peaks, which is about 360 MHz. Therefore the AOM shift is varied from 300 to 400 MHz in steps of 5 MHz. The hyperfine interval is measured by taking the unshifted spectrum from the first SAS spectrometer (using the zero-order beam from AOM1), and a shifted spectrum from the second SAS spectrometer (using the $ +1 $ order beam from AOM2).

\begin{figure}[h]
	\centering
	\includegraphics[width=0.8\linewidth]{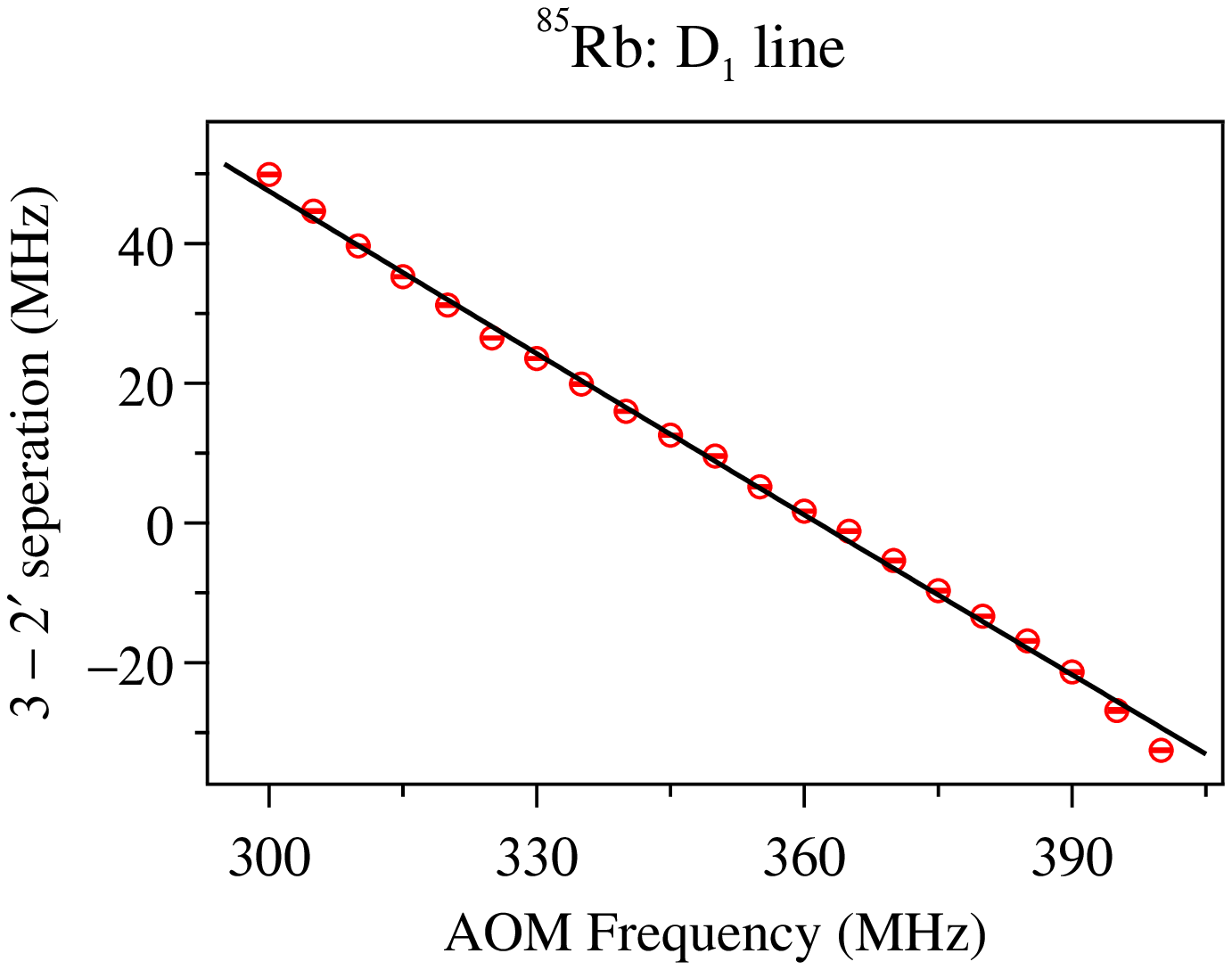}
	\caption{(Color online) Peak separation between the $ F_e = 3 $ unshifted peak and the $ F_e = 2 $ AOM-shifted peak, plotted as a function of AOM frequency. The solid line is a weighted 2nd order polynomial fit, weighted by the error bar for each point. The error bar is smaller than the symbol.}
	\label{fig:hyp85rb}
\end{figure}

The measured separation between the $ F_e = 3 $ unshifted peak and the $ F_e = 2 $ AOM-shifted peak as a function of AOM frequency is shown in Fig.~\ref{fig:hyp85rb}. The weighted second-order polynomial fit yields a zero crossing of $ 361.53 \pm 0.06 $ MHz, where the error is the statistical error in the curve fit. The interval is related to the hyperfine constant as $ 3A $. Hence the statistical error in the hyperfine constant is 20 kHz. Adding the other sources of error in quadrature as for the isotope, we get the value of measured in this work in $\rm ^{85}Rb$ as 
\[
A = 120.510 \pm 0.026 \ \ \text{MHz}
\]

\subsection{Comparison to earlier results}

\begin{figure}[h]
	\centering
	\includegraphics[width=0.8\linewidth]{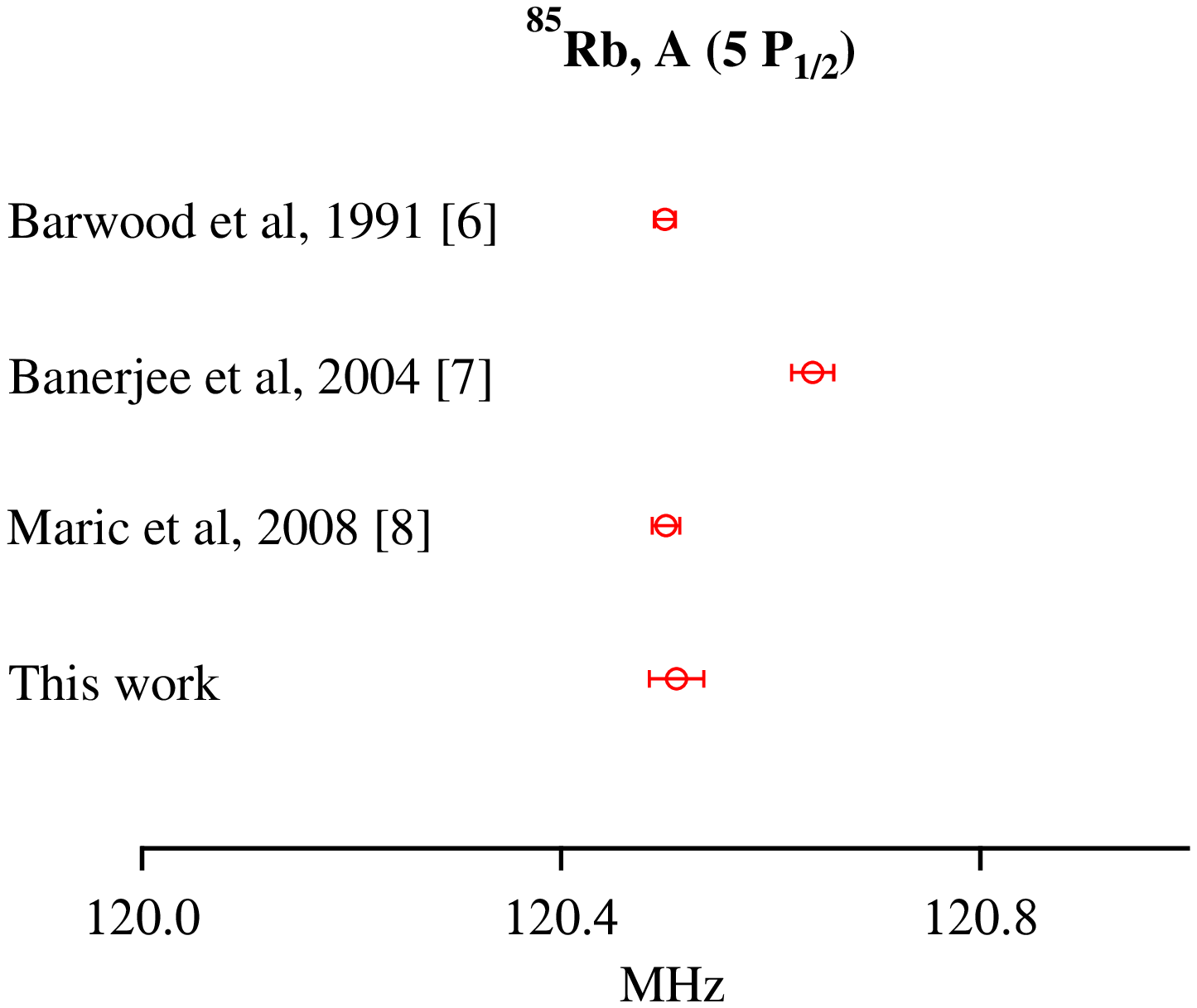}
	\caption{(Color online) Hyperfine constant $ A $ in the $ \rm 5\,P_{1/2} $ state of $ \rm {^{85}Rb} $ measured in this work compared to earlier values.}
	\label{fig:a85rb}
\end{figure}

Our result is compared to previous measurements of this hyperfine constant in Fig.~\ref{fig:a85rb}. It is seen that our present value (as for $ {\rm ^{87}Rb} $) is consistent with the values in Refs.~6 and 8, but inconsistent with a previous result from our group \cite{BDN04}.

\section{Discussion and conclusions}
In summary, we have measured hyperfine constants in the $ \rm D_1 $ line ($ \rm 5\,P_{1/2} $ state) of the two isotopes of Rb. The measurement was motivated by the fact that high-precision values reported from our lab were discrepant from the values in Ref.~\cite{BGR91}. We use a different technique from the earlier work---one in which the laser is not locked to a particular peak but scanned around it. This has the additional advantage of allowing us to verify that the lineshape of the peak is Lorentzian. The interval between two hyperfine transitions is determined by an AOM in the path of the laser beam.

After an analysis of possible systematic errors, we obtain values that have similar uncertainties to both sets of previous measurements. Our present values are inconsistent with the earlier ones from our group \cite{BDN04}, showing that this work had unaccounted systematic errors.

One possible explanation for the discrepancy is that the shift from line center when locking the laser was larger than accounted for in the error analysis. The sign of the discrepancy is also consistent with the lock point being on the higher side of the peak center, which can be understood from the fact that the photodiode signal resembles an error signal on the high-frequency side but has opposite sign on the low-frequency side. This is likely if the photodiode signal leaks through the lock in amplifier without modulation.

This explanation is reasonable because the discrepancy is about 4 MHz, which is much less than the observed linewidth of the peaks in the SAS spectrum ($\sim 20$ MHz). In fact, this was the main motivation for doing the present measurement (without locking the laser). However, this explanation is belied by the fact that our values of hyperfine constants reported in $\rm D_2 $ line of Rb (which also relied on laser locking) are consistent with previous measurements \cite{DAN08}, including ultra-precise values reported in Ref.~\cite{YSJ96}. This suggests that some other source or error must have plagued our earlier measurements.

\section*{Acknowledgments}
This work was supported by the Department of Science and Technology, India. The authors thank Pavithra Nilakandan and Sumanta Khan for help with the experiments, and S Raghuveer for help with the manuscript preparation. DD acknowledges financial support from UGC-BSR fellowship.


%

\end{document}